\documentstyle[12pt,aaspp4]{article}

\newcommand{\etal}{et al.\ }
\newcommand{\eg}{e.g.,\ }
\newcommand{\ie}{i.e.,\ }

\lefthead{von Hippel et al.}
\righthead{M35}

\begin{document}

\title{WIYN Open Cluster Study XI: WIYN\footnote{The WIYN Observatory is a
joint facility of the University of Wisconsin-Madison, Indiana University,
Yale University, and the National Optical Astronomy Observatories.} 3.5m
Deep Photometry of M35 (NGC 2168)}

\author{Ted von Hippel}
\affil{Department of Astronomy, University of Texas, C-1400, Austin, TX 78712\\
email: ted@astro.as.utexas.edu}

\author{Aaron Steinhauer}
\affil{Department of Astronomy, Indiana University, 319 Swain West, 727 East 
Third Street, Bloomington, IN 47405 \\
email: aaron@astro.indiana.edu}

\author{Ata Sarajedini}
\affil{Department of Astronomy, 211 Bryant Space Science Center, P.O. Box 
112055, University of Florida, Gainesville, FL 32611 \\
email: ata@astro.ufl.edu}

\author{Constantine P. Deliyannis}
\affil{Department of Astronomy, Indiana University, 319 Swain West, 727 East 
Third Street, Bloomington, IN 47405 \\
email: con@astro.indiana.edu}

\begin{abstract}

We present deep $BVI$ observations of the core of M35 and a nearby
comparison field obtained at the WIYN 3.5m telescope under excellent
seeing conditions.  These observations probe to $V >$ 26, and display the
lower main sequence in $BV$ and $VI$ CMDs down to $V$ = 23.3 and 24.6,
respectively.  At these faint magnitudes the background Galactic field
stars are far more numerous than the cluster stars, yet by using a
smoothing technique and CMD density distribution subtraction we are able
to recover the cluster fiducial main sequence and luminosity function to
$V = 24.6$.  We find the location of the M35 main sequence in these CMDs
to be consistent with earlier work on other open clusters, specifically
NGC 188, NGC 2420, and NGC 2477.  We compare these open cluster fiducial
sequences to stellar models by Baraffe \etal (1998), Siess \etal (2000),
Girardi \etal (2000), and Yi \etal (2001) and find that the models are too
blue in both $B-V$ and $V-I$ for stars less massive than $\sim0.4$
M$_{\odot}$.  At least part of the problem appears to be underestimated
opacity in the bluer bandpasses, with the amount of missing opacity
increasing toward the blue.  M35 contains stars to the limit of the
extracted main sequence, at M $\approx$ 0.10--0.15 M$_{\odot}$, suggesting
that M35 may harbor a large number of brown dwarfs, which should be easy
targets for sensitive near-IR instrumentation on 8--10m telescopes.  We
also identify a new candidate white dwarf in M35 at $V = 21.36 \pm
0.01$.  Depending on which WD models are used in interpreting this
cluster candidate, it is either a very high mass WD ($1.05 \pm 0.05$
M$_{\odot}$) somewhat older (0.19--0.26 Gyr, 3--$4\sigma$) than our best
isochrone age (150 Myr), or it is a modestly massive WD (0.67--0.78
M$_{\odot}$) much too old (0.42--0.83 Gyr) to belong to the cluster.
Follow-up spectroscopy is required to resolve this issue.

\end{abstract}


\section{Introduction}

NGC 2168 (M35) is a rich open cluster with an age similar to the
Pleiades.  Since M35 is more populous and covers a smaller angular extent
than the Pleiades, it offers excellent opportunities for studies of
stellar evolution at $\sim100$ Myr, even though M35 is further away and
suffers greater background contamination.  Astrometric studies of M35 have
a long history (Ebbighausen 1942; Meurers \& Schwarz 1960; Lavdovskij
1961; Cudworth 1971; McNamara \& Sekiguchi 1986a), and in fact continue to
this day--M35 is {\it the} astrometric calibrator for the HST Fine
Guidance Sensors (McArthur \etal 1997). Modern photometric studies of this
cluster begin with Sung \& Lee (1992) who obtained photoelectric $UBV$
photometry for 112 field plus cluster stars to $V = 14$, approximately the
same limiting magnitude as the two more recent proper motion studies.
Sung \& Lee derived a true distance modulus of 9.3, an age of 85 Myr, and
internal differential reddening of $0.26 \leq$ E($B-V$) $\leq 0.44$.  In a
subsequent study, Sung \& Bessel (1999) obtained $UBVI$ CCD photometry for
stars brighter than $V = 20$ in a central $20^{\arcmin}.5 \times
20^{\arcmin}.5$ cluster field.  From these data they derived
$(V-M_{V})_{\rm o} = 9.6 \pm 0.1$, E$(B-V) = 0.255 \pm 0.024$
(corresponding to $(V-M_{V}) = 10.39$ for $R_{V} = 3.1$), log age = 8.3
$\pm$ 0.3 (200 Myr), [Fe/H] $\approx -0.3$ (based on $U-B$ color excess),
a present day mass function slope of $-2.1 \pm 0.3$,\footnote{All mass
function slopes presented here are on the system $n(m) \propto
m^{-(1+x)}$, where the reported slope = $-(1+x)$, and the Salpeter (1955)
value is $-2.35$.} and a binary frequency $\geq 35$\%.  A younger cluster
age of 70 to 100 Myr was found by Reimers \& Koester (1988a), based on a
reanalysis of older photometry along with isochrones from Maeder \&
Mermilliod (1981) and on the cooling age of two cluster white dwarfs.
Barrado y Navascu\'es, Deliyannis, \& Stauffer (2001a) have derived the
cluster metallicity, [Fe/H] = $-0.21 \pm 0.10$, from high resolution
spectroscopy.  In the most recent photometric study of M35, Barrado y
Navascu\'es \etal (2001b, hereafter BSBM), using the Kitt Peak 4m and CFH
3.6m telescopes, imaged the central $28{\arcmin} \times 28{\arcmin}$ of
the cluster in $VRI$ to $V \approx 22$ and $I \approx 23$.  BSBM found a
luminosity function similar to the Pleiades, with a peak near $M_{I} = 9$,
and present-day mass function characterized by three different power law
slopes over the mass range 1.6 to $\sim0.1$ M$_{\odot}$.  BSBM found their
central cluster field to contain $\sim1600$ M$_{\odot}$ among cluster
members.

In a dynamical study of M35, Leonard \& Merritt (1989) found that M35 is
close to dynamical equilibrium, that its dynamical mass within the central
3.75 pc is 1600 to 3200 M$_{\odot}$ (95\% confidence), and that its IMF
slope is $-2.7 \pm 0.4$ between 1 and 6 M$_{\odot}$.  Mathieu (1983),
McNamara \& Sekiguchi (1986b), and BSBM all noted that M35 exhibits mass
segregation, though it is unclear whether this is due to relaxation or
initial conditions.  Mathieu (1983) pointed out that the cluster age is
close to the expected relaxation time of the intermediate mass component,
though the relaxation time scale is uncertain by a factor of 2.

We obtained deep $BVI$ photometry of M35 in order to study the low mass
main sequence stars and to search for cluster white dwarfs.  Our study
presents higher signal-to-noise data for the faintest stars than the BSBM
study, and we achieve this depth in $B, V$, and $I$, whereas BSBM achieve
their greatest depth in $R$ and $I$.  Our photometry allows us to
isolate the fiducial main sequence of the cluster in $B$ and $V$, as well
as $V$ and $I$, which we compare to a range of stellar models.  The
smaller field of view of our study precludes a detailed luminosity
function or mass function study, however, as done by BSBM.  In these
trade-offs between field of view and depth in various filters, our two
deep photometric studies are complementary.  In addition, we have found a
candidate cluster white dwarf which, if a {\it bona fide} cluster member,
places constraints on a combination of cluster age and stellar evolution.

\section{Data Reduction}

We observed M35 and a nearby comparison field at the WIYN 3.5m telescope
located on Kitt Peak through Harris $B$- and $V$-band and Mould
interference $I$-band filters (Massey \etal 1987) on the nights of
December 31, 1997, January 1, 1998, and January 22, 1998.  All of these
nights were non-photometric, though the seeing was excellent, ranging from
0.46 to 1.0 arcseconds FWHM, with the majority of the observations
obtained during 0.6 to 0.8 arcsecond seeing conditions, even in the
$B$-band.  Total exposure times of 4500, 3357, and 7200 seconds in $B$,
$V$, and $I$, respectively, were obtained on a field centered on M35, at
RA = $06^{h}08^{m}54^{s}$, Dec = $+24{\arcdeg}17{\arcmin}53{\arcsec}$
(2000), corresponding to Galactic coordinates {\it l, b} = 186.62, 2.17
degrees.  Total exposure times of 4500, 4500, and 9000 seconds in $B$,
$V$, and $I$, respectively, were obtained in a nearby comparison field,
located at RA = $06^{h}08^{m}54^{s}$, Dec =
$+24{\arcdeg}33{\arcmin}59{\arcsec}$ (2000), corresponding to {\it l, b} =
186.39, 2.30 degrees.

The CCD detector then in use at WIYN, ``S2KB,'' is a $2048^2$ STIS CCD
with 21 micron (=0.2{\arcsec}) pixels and a field of view of $6.8 \times
6.8$ arc minutes.  Data reduction followed standard procedures virtually
identical to those described for the old open cluster NGC 188 by von
Hippel \& Sarajedini (1998, hereafter WOCS1).  Briefly, we removed a
time-dependent, two-dimensional bias structure with the help of the
overscan regions and standard bias frames to within a typical accuracy of
1 ADU.  For these broad-band exposures the high sky meant that this bias
uncertainty was always much less than 1\%.  Flat fielding was performed
using dome flats with typical pixel-to-pixel Poisson uncertainties of $\le
0.25$\% and illumination pattern uncertainties $\la 1.0$\%.

Instrumental magnitudes were extracted from each image individually using
ALLFRAME (Stetson 1994). We employed a quadratically varying PSF, defined
using the brightest 50--100 isolated stars, and refined by iteratively
subtracting faint neighbors. These PSFs along with a master coordinate
list of stellar positions for each field were input into ALLFRAME which
then produced total PSF magnitudes for all detected profiles. The PSF
magnitudes measured for each frame were corrected to total magnitudes by
applying a spatially variable aperture correction. The amplitude of this
correction was never more than $\pm$0.02 mag but was nevertheless applied
to maximize the accuracy of the resultant photometry.  We edited the
photometry using fitting quality diagnostics provided by ALLFRAME. In
particular, we set a maximum overall CHI value of 2.5 and a maximum mean
magnitude error in 0.5 mag bins of 2$\sigma$. We also stipulated that the
distribution of SHARP values be symmetric around zero and thus edited
outlying SHARP values until this was accomplished.

Since the nights were not photometric, the WIYN data were placed on the
standard system of Landolt (1983, 1992) using photometric, calibrated
$BVI$ observations (Deliyannis \etal 2002) of the central $23{\arcmin}
\times 23{\arcmin}$ of M35 taken on the night of October 24, 1998 with the
KPNO 0.9m telescope.  The 0.9m M35 observations were first calibrated via
the Landolt standards, and then the 0.9m data were used to calibrate the
WIYN data.  The details of the standards observed to calibrate the 0.9m
M35 observations are presented in Table 1.  Column 1 lists the filter in
question, with two different color calibrations given for the $V$-band,
one based on the $B-V$ color and the other based on the $V-I$ color.
Column 2 lists the number of independent CCD frames taken of the standard
fields, column 3 lists the number of standards used to derive the
photometric transformations, columns 4 and 5 list the photometric zero
points and their associated errors, columns 6 and 7 list the derived
airmass coefficients and their errors, and columns 8 and 9 list the
derived color terms and their errors.  The standard observations spanned
the same UT range as the cluster observations, though not the same airmass
range.  The airmass range for the standards was 1.18--1.64, whereas the
cluster was observed over a lower and more restricted airmass range,
1.01--1.06.  While the Landolt standards were observed at a slightly
higher minimum airmass than M35, this was unavoidable given our desire to
observe M35 at the lowest possible airmass, whereas the Landolt fields are
located on the celestial equator.  The airmass terms were linear, so this
extrapolation to airmass = 1.0 (from airmass = 1.18) should have a minimal
effect on the standardization.  In our standardization procedure we define
the color terms for all filters with respect to $V$.  For the $V$
calibration, $B-V$ was used preferentially.  However, the $V-I$ color term
was also calculated for $V$, and used when a $B$ magnitude was not
available for a given program star.  This allowed us to take full
advantage of our greater photometric depth in $V$ and $I$ than in $B$.
The color terms were linear, small, and had tiny uncertainties.

While the 0.9m M35 observations probed to $BVI$ = 20--21, $\sim 5$ mag
shallower than the WIYN observations, there remained $> 4$ magnitudes of
overlap between the two data sets, and the stars in common cover much of
the range in color (for $B$ the range was $B-V$ = 0.92--1.89 and for $VI$
the range was $V-I$ = 0.17--2.28).  We thus calibrated the WIYN photometry
from the many (92 to 292, depending on the filter and field) stars in
common with the 0.9m photometry.  We calculated transformations between
the two data sets, taking into account possible changes in the color
dependence of the offset.  For the $B$ filter, the following equations
were solved:

\begin{equation}
(B-V) = a_1\ (b-v) + c_1
\end{equation}
\begin{equation}
b-B = a_2\ (B-V) + c_2
\end{equation}

where $b$ and $v$ are instrumental magnitudes and $B$ and $V$ are standard
magnitudes.  The constants were empirically determined.  Similar equations
were used to standardize the $V$ and $I$ data using the $V-I$ color in
the transformation.

Based on experience with these instruments and similar data, we estimate
the total systematic error of our standardized photometry to be $\la 0.02$
mag.  Note, however, that the reddest Landolt standard we used had $B-V$ =
1.9 and $V-I$ = 2.4, whereas our data continue to $B-V$ = 2.4 and $V-I$ =
4.0.  Since the color terms for the lower main sequence are
extrapolations, the photometry for the reddest stars is likely to be less
accurate than the numbers quoted above.  Since the reddest stars are of
particular importance in this study we explored two avenues to constrain
any possible systematic shifts in the photometry.  In the first approach,
we perturbed all the photometric transformation coefficients by $+1\sigma$
and by $-1\sigma$.  This approach only indicates whether errors in the
transformation coefficients themselves matter in the extrapolation, of
course, and not whether different transformation coefficients are
appropriate for the reddest stars.  The mutual probability that all four
(per filter) transformation coefficients would be each wrong by $+1\sigma$
or $-1\sigma$ is $< 5$\% (\ie a $2\sigma$ error).  This $2\sigma$ envelope
at the very bottom of the main sequence corresponds to only $\Delta(B-V) =
\pm 0.09$ and $\Delta(V-I) = \pm 0.014$.  This small error range is
gratifying and due to the fact that the color terms are small and linear,
and the photometric calibration, especially in $V$ and $I$, was
excellent.  As a second constraint on any systematic errors in the color
terms we performed synthetic photometry for a range of F0V to M6V stars
from the Pickles (1998) spectral library, using the tracings for the WIYN
filters and the measured quantum efficiency versus wavelength for the S2KB
CCD.  The synthetic photometry was compared to standard $B-V$ and $V-I$
colors for stars of the same MK type compiled by Johnson (1966).  Overall
we found excellent agreement between our synthetic color terms and those
we derive from the standard star observations, except for the very reddest
stars where our synthetic photometry indicated that our extrapolated color
terms may make the faintest cluster members too blue in both $B-V$ and
$V-I$ by 0.1 to 0.2 mag.  We do not apply the offset indicated by the
synthetic photometry, but rather use it as a guideline to the level of the
likely systematic color error at the limit of the main sequence.  This
level of error is acceptable and less than the differences we will explore
between the data and stellar models.

As a final step in the data reduction process, we measure image morphology
using SExtractor (Bertin \& Arnouts 1996) in order to reject non-stellar
objects.  SExtractor has the advantage of determining sky locally and uses
verified neural network techniques to perform the star/galaxy
classification.  Figure 1 shows the results of the SExtractor
morphological classification versus $I$-band magnitude for both the M35
and control fields.  The morphological ``Stellarity Index'' ranges from 0
for galaxies or obvious imaging defects to 1 for stars.  Although
SExtractor's neural network classifier is not strictly a Bayesian
classifier, the Stellarity Index values are approximately the
probabilities that the object is a point source.  The classifications near
0.5 at the faintest magnitudes demonstrates the common sense notion that
at limiting signal-to-noise classification breaks down.  Objects from the
M35 central field are plotted as small filled circles.  Objects from the
control field are plotted as small open circles and shifted down by 0.1
and to the right by 1 mag, so that both data sets can be clearly seen in
the same plot.  Figure 1 demonstrates that the morphological-magnitude
distribution of objects in both fields is essentially identical, the
result of observations obtained under excellent and stable conditions.
The large number of definite stars in Figure 1 is the result of a cluster
plus a rich Galactic field, and the significant suppression of background
galaxies is caused by the high line-of-sight absorption in front of and
behind M35.  Still, there is extragalactic contamination, which we now
remove.

We select those objects determined by SExtractor to have a high ($\geq
0.95$, marked by the dashed lines in Figure 1) probability of being stars
based on their $I$-band morphology, since the $I$-band is the deepest and
best resolved, particularly for the faintest red cluster stars.  While
some of the objects with lower stellarity indices may be stars, we wish to
be conservative in our attribution of objects to the cluster, especially
at the faintest magnitudes and intermediate colors, where galaxies may
contribute significantly to the number counts.  As it turns out, there are
few contaminating galaxies brighter than $I = 22$, above which we have
excellent morphological classifications.  For $I \leq 22$, we reject only
3.5\% and 1.2\% of the objects in the central and control field as being
non-stellar, respectively.  Fainter than this limit the morphological
classifications begin to degrade, but this is so faint, particularly on
the main sequence where it is equivalent to $V = 26$, that it has no effect
on any of the statistical cluster quantities that we derive.

The cleaned and calibrated color magnitude diagrams (CMDs) are presented
in Figures 2a through 2d.  In these figures the morphological criterion is
essentially equivalent to a signal-to-noise cut (at $\sim15$) in the
$I$-band.  Since the $B$- and $V$-band observations are shallower than the
$I$-band observations, however, the limiting depths in Figures 2a-d are
not due to our morphology statistic.  Individual error bars are not
plotted for clarity but typical photometric errors at each integer
$V$-band magnitude starting at $V = 16$ are presented down the right-hand
sides of Figures 2a-d.  These errors are internal errors only.

\section{Discussion}

The dynamic range of our WIYN observations is approximately 10 magnitudes,
with the cluster main sequence evident over 8 or 9 magnitudes.  For
comparison, note that the cluster turn-off is another 7 magnitudes
brighter than the brightest photometry presented here, at $V \approx 8$.
Although M35 is a rich cluster, as is particularly evident along the
well-populated upper main sequence (see figure 3 of Sung \& Bessell 1999)
or from wide field images such as the Palomar Observatory Sky Survey, when
viewed in deep CMDs as presented here, there are clearly far fewer stars
belonging to M35 than to the background Galactic field.  This is expected
given the low Galactic latitude and the tremendous volume surveyed behind
the cluster.  Typical field stars are 1 to 4 mags fainter than the cluster
main sequence at a given $B-V$ or $V-I$ color.  Even though most of these
objects are main sequence stars, they still reside at distances of 1.6 to
6.3 times further away than the cluster, and thus these CMDs survey a huge
volume of the background Galactic disk.

Even with the copious background, the cluster main sequence is discernible
in the central field, and possibly in the control field.  At
$16{\arcmin}.1$ north of the cluster center, our control field may not
completely avoid cluster stars.  This will be discussed in a statistical
context, below.  (In fact, the control field was initially selected to be
within the cluster periphery so that we could measure luminosity functions
in both the cluster center and periphery in order to look for signatures
of mass segregation.  The low number density of faint cluster stars
precluded that comparison, but left us with a very useful control field
that is a good compromise between avoiding the cluster entirely and moving
so far away that the background field is not comparable--a potential
problem given the steep gradient in stellar counts near the Galactic plane
and the variable, though not large, reddening in this field.)  Visually,
the cluster main sequence can be traced to $V \geq 25$ in the $V-I$ CMD.
The cluster binary sequence is not obvious in these CMDs (more on this
later).  Finally, a few blue objects between $V = 21$ and $V = 22$ can be
seen in both the central and control field CMDs.  We explore the
likelihood that any of these objects are cluster white dwarfs and the
implications for the cluster and white dwarf physics below.

\subsection{Cluster Parameters}

We adopt the cluster parameters from our wide-field study in preparation
(Deliyannis \etal 2002), \ie $(m-M)_{V} = 10.15$, E($B-V$) = 0.20, and age
= 150 Myr, although we will check the validity of the distance and
reddening assumption based on fitting the 0.9m main sequence photometry to
other cluster main sequences observed by Hipparcos (Pinsonneault \etal
1998).  We also adopt E($V-I$) = 1.34 E($B-V$) based on the central
wavelengths of the WIYN $V$ and $I$ filters and the relations of Cardelli,
Clayton, \& Mathis (1989).  Note that our distance and reddening are
roughly consistent with, but not identical to, those used by BSBM, who
adopted E($B-V$) = 0.255 from Sung \& Bessell (1999) and who used an
$I$-band distance modulus corresponding to $(m-M)_{V} = 10.37 \pm 0.1$.
Before performing the main sequence fitting, we first extract the cluster
fiducial sequence by employing our control field to remove the
contaminating Galactic field stars.

\subsection{Removing Field Star Contamination}

The M35 and comparison fields were observed to similar depths under
similar conditions, presenting us with sufficient data to perform a good
statistical subtraction of the contaminating field stars.  As a first
attempt at this subtraction we identified nearest neighbors in the
central and control CMDs, then subtracted them.  We also tried variants on
this technique where the number of objects in boxes of different sizes
were computed in each CMD, then subtracted.  This entire approach proved
unsatisfactory and it became clear that an appropriate form of smoothing
was required.  If each CMD is thought of as an estimated density
distribution, then the difference between a cluster field CMD and a
control field CMD becomes the difference between two estimated density
distributions.  As advised by Silverman (1986), we first smoothed the CMDs
with an Epanechnikov kernel.  The Epanechnikov kernel is an inverted
parabola that places the peak density at the object position and zero
density at plus and minus the kernel width.  This smoothing was performed
using software kindly provided by K. Gebhardt (for examples of its use in
astronomy see Gebhardt \etal 1995, 1996).  We tried a variety of
Epanechnikov kernel sizes (smoothing lengths), ranging from 0.01 mag to
1.0 mag and found that 0.10 mag represented the best compromise between
maintaining the original resolution of the CMDs and smoothing the CMDs
enough so that the subtraction did not contain too much high frequency
noise.

A slight complication in the CMD subtraction was that the comparison field
is slightly less reddened than the cluster field.  This difference was
unsurprising given the known variable reddening in the vicinity of the
cluster.  The decreased reddening in the comparison field relative to
the central field was measured by fitting the blue edge of the field
stars in the CMD in three different magnitude ranges.  We estimate the
reddening difference to be $\Delta$E($B-V$) = 0.05.  Although this
reddening difference is almost certainly spread out in distance along the
line of sight, we approximate it as a single value for all stars in the
comparison field, and apply a single offset in color and luminosity to the
control field CMDs.  Small errors in this reddening offset and the
limitations of using a single reddening value will make our subtraction
process noisier, but will not meaningfully affect the location of the
derived fiducial cluster sequences.  Note that alternative explanations
for the color offset between the cluster and comparison fields, \eg due to
stellar population differences, are unlikely.  These two fields are near
enough to one other that the Galactic model of Reid \& Majewski (1993)
yielded essentially identical CMDs for both fields.  Additionally, the
$VI$ CMD for the comparison field had slightly more stars than its
counterpart for the cluster, 1554 stars versus 1463 stars, despite the
fact that the comparison field is at a slightly higher Galactic latitude,
$b$ = 2.30 versus $b$ = 2.17.  The excess is primarily at the bottom of the
CMD, past $V = 25.3$, consistent with expectations of less absorption for
this field.

Figure 3a presents a smoothed (kernel = 0.10 mag) $VI$ CMD for the central
cluster field and Figure 3b presents a similar $VI$ CMD for the control
field, artificially reddened by E($B-V$) = 0.05.  Figure 3c presents the
subtracted CMD.  For Figures 3a through 3c the horizontal axes are $V-I$
color from 0.0 to 4.5 and the vertical axes are $V$ magnitude from 15 to
26.  Figure 3c also presents the best fit fiducial sequence as a white
line.  The fiducial sequence has been shifted downward by 0.30 mag for
presentation purposes, to make it easier to visually identify the cluster
main sequence.  While there is some subtraction noise elsewhere in the
CMD, seen as black points which are negative star counts, the cluster main
sequence is almost entirely positive.  Due to the limited statistical
significance of the presence of the main sequence above $V = 17.7$ in these
WIYN data, the fiducial sequence above that limit was derived from the
wide-field 0.9m data (Deliyannis \etal 2002), and checked for consistency
in these data.  We created similar smoothed and subtracted $BV$ CMDs.

\subsection{Fiducial Main Sequence and Distance}

We extracted the cluster fiducial main sequence from the $VI$ (Figure 3c)
and $BV$ subtracted density maps by identifying peaks in the main sequence
region. Because of the small number of stars involved we found it easiest
to first identify the main sequence by eye, then statistically validate
and improve our choice of main sequence location by computing star counts
in regions around the nominal fiducial sequence.  Star counts determined
in regions around the correct main sequence location should asymptote to a
constant value after a sensible extraction width has been reached since
only in the main sequence region should there be excess star counts in the
difference maps.  To demonstrate this point, the solid curve cutting
through the circles in Figure 4 presents the cumulative number of stars
extracted along the entire $V$-band magnitude range (19.82 $\leq V \leq$
24.62) of our initial WIYN $VI$ fiducial sequence as a function of the
width of the extraction window.  The smallest extraction window considered
was $\pm 0.00$ mag, \ie along the 0.01 width of this fiducial sequence.
The largest window considered was $\pm 0.5$ mag, for a total width of 1.01
mag.  Once the window width reached 0.29 mag ($\pm 0.14$ mag around the
fiducial sequence) the total number of stars in the LF remained constant,
at 47.  The quality of the subtraction around the main sequence region
remains good even with a window width as large as $\sim0.7$ ($\pm 0.35$)
mag.  Besides the curve indicating cumulative star counts around the
initial fiducial sequence as a function of $V-I$ window width, a series of
other curves are plotted, each for a different fiducial sequence created
by offsetting the initial fiducial sequence by $\pm 0.025$, 0.05, 0.075,
0.10, and 0.125 mags.  The $+0.025$ mag offset (redward) represents a
slight improvement over our initial fiducial sequence, and we adopt this
improvement for our best estimate of M35's fiducial sequence.  Figure 4
also serves to yield an overall $V-I$ error in the fiducial sequence:  The
five curves containing the most stars within the smallest color range
present the reasonable range of the best fit, yielding an overall
uncertainty in the fiducial sequence of $\Delta(V-I) = \pm0.05$.

The identical procedure was used for the $BV$ cluster fiducial, where we
found a color uncertainty, $\Delta(B-V) = \pm0.03$. While our procedure
does not justify the subtle wiggles in the $BV$ and $VI$ fiducial
sequences, it does justify the overall fiducial location, particularly for
$V > 22.5$, where the bulk of the cluster stars are found, and where we
will focus our comparison with stellar models in the next section.

For presentation purposes and as a consistency check on the cluster
distance, we extended the WIYN fiducial sequences to brighter stars ($V
\leq 16.52$, $B-V \leq 1.1$ and $V \leq 17.5$, $V-I \leq 1.5$), using the
0.9m photometry of Deliyannis \etal (2002).  The bright fiducial sequences
are estimated by eye without the aid of the CMD subtraction technique
discussed above.  We estimate the errors in the bright fiducial fits to be
$\pm 0.10$--0.15 mag in $V$ for a given $B-V$ or $V-I$ color, for $V \geq
17$.  In the overlap region, between $V$ = 17 and 20, where the 0.9m data
are not as precise and where the WIYN data contains fewer cluster stars,
the error is approximately $\pm 0.20$ mag.  The $BV$ and $VI$ fiducial
sequences are presented in Tables 2 and 3, respectively.  Note that these
fiducial sequences are presented in the observed system.  To convert to
the absolute system, $M_V$ and $(B-V)_{\rm o}$ or $(V-I)_{\rm o}$, the
user can apply our preferred distance and reddening ($(m-M)_{V} = 10.15$,
E($B-V$) = 0.20) or their own, along with E($V-I$) = 1.34 E($B-V$),
appropriate for these filters.

The upper part of the fiducial sequence, assuming our adopted cluster
parameters, is compared in Figure 5 to the Hipparcos cluster fiducial main
sequence derived by Pinsonneault \etal (1998).  The Hipparcos fiducial is
based on five nearby open clusters ($\alpha$ Per, Coma Ber, Hyades,
Pleiades, and Praesepe) and valid over the range $0.55 \leq (V-I)_{\rm o}
\leq 0.9$.  Although Pinsonneault \etal find a systematic problem with the
Pleiades parallaxes, they argue that the other four open clusters are
consistent with the same fiducial sequence within very stringent limits.
They conclude that with good data one can derive the distance to a near
solar metallicity open cluster using the main-sequence fitting technique
to an accuracy in the distance modulus of $0.05$ mag.  The correction away
from solar metallicity to the cluster metallicity, [Fe/H] = $-0.21 \pm
0.10$ (Barrado y Navascu\'es \etal 2001a), would require a shift in
distance modulus of $-0.13 \pm 0.06$ mag (Pinsonneault \etal 1998).
Besides this possible systematic shift due to metallicity, the fit between
our cluster fiducial sequence and the Pinsonneault \etal fiducial is
excellent, with a residual of $-0.033 \pm 0.033$ mag (fitting error) $\pm
0.087$ mag (error in fiducial points) = $-0.033 \pm 0.093$ if we use the
three M35 fiducial points within the Pinsonneault \etal fitting range.
Expanding the fit to the two points immediately outside this fitting
range, and giving these points half the weight of the central three
points, we find a residual of $-0.051 \pm 0.027 \pm 0.075$ mag = $-0.051
\pm 0.080$.  Main sequence fitting of the Pinsonneault \etal sequence to
our 0.9m data for M35 supports our cluster distance modulus within $\pm
0.09$ mag, with perhaps marginal evidence that the cluster distance
modulus is slightly greater than the one we employ, $(m-M)_{V} = 10.15 +
(0.03$ to 0.05 main sequence offset) + ($-0.13 \pm 0.06$ possible
metallicity offset).

Before proceeding to compare M35's fiducial main sequence to stellar
evolution models we first compare it to the fiducial main sequence
presented by BSBM, which they derived primarily by fitting a sequence to
field star photometry with accurate parallaxes presented by Leggett
(1992), and secondarily, for the bright stars, by fitting the photometry
of Sung \& Bessell (1999).  This comparison is plotted in Figure 6, where
we use our adopted distance modulus and reddening to convert both observed
fiducial sequences to a common absolute magnitude and dereddened $V-I$.
Our fiducial sequence is $\sim0.05$ mag redder in $V-I$ for $M_{V} \geq
8$, and significantly redder at the faintest magnitudes, corresponding to
a shift of 0.20 mag at $M_{V} = 14$.  Alternatively, our fiducial sequence
is brighter than BSBM's fiducial sequence, with a systematically
increasing luminosity difference for the faintest stars.  It is unclear
what the source of this color or luminosity offset could be.  The BSBM
fiducial is drawn from a heterogeneous local field stellar sample, so
perhaps this heterogeneity is the cause of the difference.  Yet on average
their field stars should have the same or greater metallicity than M35,
which would produce an offset in the opposite direction.  Also, as
discussed earlier, any systematic due to extrapolating our color terms for
the reddest stars would not help, since the expected correction would
only make our faintest main sequence stars redder.

In Figure 7 we compare the fiducial main sequences of M35, NGC 188
(WOCS1), NGC 2420 (von Hippel \& Gilmore 2000), and NGC 2477 (derived
from von Hippel \etal 1996).  The latter two cluster sequences are based
on HST photometry, whereas the former two are based on WIYN photometry,
though the NGC 188 data were directly calibrated at WIYN while these M35
data were calibrated with 0.9m photometry.  All four clusters have the
potential for systematic photometry errors among the reddest stars, due to
the same problem with insufficient standards for $V-I \ga$ 2.  The HST
calibrations, for instance, degrade past $V-I = 1.5$ (Holtzman \etal
1995).  Hopefully, however, these systematics manifest themselves
differently in the different instruments and the range in the photometry
is an indication of their reliability.  In addition, there are slight
differences in metallicity, with [Fe/H](NGC 2420) $\approx -0.4$,
[Fe/H](M35) $\approx -0.2$, and [Fe/H](NGC 188) $\approx$ [Fe/H](NGC 2477)
$\approx 0.0$.  These small metallicity differences only correspond to
small color or luminosity shifts, with a deficiency of 0.2 dex in [Fe/H]
corresponding to a shift blueward by $B-V \approx 0.005$--0.04 and $V-I
\approx 0.04$--0.07, depending on the luminosity.  After incorporating the
effects of metallicity, the lower main sequence of M35 appears to be
redder than the other clusters by $\Delta(V-I) = 0.2$ mag.  Although this
offset is most likely due to the uncertainties in the color
transformations, we do not correct for it, since we don't know which
clusters are in error.  Fortunately, the differences in sequences are
minor compared to the current uncertainties in the stellar evolution
models.

\subsection{Comparison to Stellar Models}

We now compare our fiducial sequences to stellar models by four groups.
Readers should keep in mind that all groups acknowledge difficulties
matching observations below $\sim0.4$ M$_{\odot}$ (corresponding to
$T_{\rm eff} \approx 4000$ K, $B-V \approx 1.3$, and $V-I \approx 2.0$).
The reasons for difficulties with the cool stars are many: the fundamental
problem of convection, uncertainties in the opacities, the equation of
state, and creating reliable model atmospheres.  For instance, the
incomplete treatment of convection and uncertainties in the opacities,
particularly molecular opacities, affect model radii.  Without reliable
model radii, it becomes impossible to create reliable color-$T_{\rm eff}$
transformations.  With these difficulties in mind, our comparisons are
meant to help quantify the level of mismatch between theory and
observation in two common ($BV$ and $VI$) broad-band CMDs, and to help
observers choose appropriate models and estimate errors when studying
these low mass stars.

Figure 8a compares the $VI$ cluster fiducials for M35 and NGC 188 to solar
metallicity stellar isochrones from Baraffe \etal (1998).  The small
subsolar metallicity for M35 corresponds to a shift in the models which
makes them nearly match M35 above 0.45 M$_{\odot}$, but exacerbates the
difference between the models and the clusters below 0.45 M$_{\odot}$.
The 158 Myr and the 6.3 Gyr models are the models closest in age to M35
(150 Myr, Deliyannis \etal 2002) and NGC 188 ($7 \pm 0.7$ Gyr, Sarajedini
\etal 1999), respectively.  Mass values in solar units are indicated along
the 158 Myr isochrone.  According to these models, our M35 photometry
extends down to $\sim0.10$ M$_{\odot}$.  Both the 158 Myr and 6.3 Gyr
isochrones are too blue for stars less massive than $\sim0.45$
M$_{\odot}$.  The slope of the fiducial main sequences and the model
isochrones begin to diverge somewhat earlier, between $M_{V}$ = 8 and 9.
In Figure 8b updated, unpublished models kindly supplied by I. Baraffe
are compared to the same data.  These unpublished models are identical to
those of Baraffe \etal (1998) except that the TiO line list from Schwenke
(1998) instead of J\/orgensen (1994) was used (see also section 3.1 of
Chabrier et al. 2001 for details).  These models appear to be an
improvement: the color discrepancy with the lowest mass stars is reduced
by a factor of approximately two.

Figure 9a compares the $BV$ M35 fiducial sequence to [Fe/H] = $-0.3$
stellar isochrones from Siess, Dufour, \& Forestini (2000).  Two separate
color transformations for the isochrone appropriate to M35 are plotted,
one compiled by Siess, Forestini, \& Dougados (1997) and the other from
Kenyon \& Hartmann (1995).  Figure 9b provides the same comparison in the
$VI$ CMD, this time also including NGC 188 and an appropriate isochrone.
To approximately the same degree as the updated Baraffe \etal models, the
Siess \etal models are too blue starting at $M_{V}$ = 8--9, and we see
that the problem is not limited to $VI$ photometry.

Figures 10a and 10b provide similar comparisons to the Girardi \etal
(2000) isochrones, for which the deviation between the models and cluster
fiducials are more pronounced, especially in the $VI$ CMD.  For the last
model comparison, Figures 11a and 11b compare the cluster fiducials to
Yonsei-Yale (Yi \etal 2001) isochrones.  Since Yi \etal provide an
interpolation tool with their isochrones, we created a 150 Myr, [Fe/H] =
$-0.2$ isochrone to precisely match M35's parameters, and a 7 Gyr, [Fe/H]
= 0 isochrone to precisely match NGC 188's parameters.  Two different
color calibrations are presented--one (indicated by a solid line) by
Lejeune, Cuisinier, \& Buser (1998) and the other (indicated by a dotted
line) an updated version of the transformations of Green, Demarque, \&
King (1987).  The Yonsei-Yale models are also too blue for much of the
lower main sequence in the $BV$ CMD, but surprisingly are too red in the
$VI$ CMD, for $V-I \ga 2.8$.  The constant slope at the lower mass end,
however, is an indication that these models are extrapolating at least
some of the physics applicable for intermediate mass stars into the low
mass regime, in this case the equation of state is extrapolated below 0.45
M$_{\odot}$ (P.  Demarque, private communication).  To be fair, the
Girardi \etal and the Yi \etal models were not designed for this very low
mass range, and we present them along with the Baraffe \etal and Siess
\etal models, which were designed for this mass range, only to present a
more complete comparison with current stellar models.

In essentially all cluster-to-model comparisons the models are too blue in
both $B-V$ and $V-I$ for stars less massive than $\sim0.4$ M$_{\odot}$.
It is unlikely that errors in our photometric calibration cause this
discrepancy since the lower main sequence we derive for M35 is consistent
with the lower main sequences for three other star clusters and since the
likely systematic error in our color terms would only make the actual
stars 0.1 to 0.2 mag redder, increasing the discrepancies with the
models.  At least part of the cluster-to-model mismatch appears to be
underestimated opacity in the bluer bandpasses, with the amount of missing
opacity increasing toward the blue ($B - V$ already includes the excess
$V$-band opacity apparent in $V - I$).  More generally, at low
temperatures, both the model radii and the color transformations between
the theoretical $L$-$T_{\rm eff}$ plane and the observational plane are
still uncertain, due to the unphysical parameterization of convection
(mixing length theory) and deficiencies in current model atmospheres
(particularly molecular transition data), respectively.  We look forward
to developments in these theories and more sophisticated models that fully
treat radiative hydrodynamics.

\subsection{Luminosity Function}

The steps used to derive the $VI$ fiducial sequence from the subtracted
$VI$ CMD density map were essentially the same as deriving the cluster
luminosity function.  The process of shifting the comparison field
slightly to correct for its lower reddening, smoothing the CMDs,
subtracting the smoothed control field CMD from the cluster field CMD, and
identifying and statistically validating the location of the cluster main
sequence, leads us to the point where we need only extract the number of
stars in the differenced CMD along the main sequence.  The only parameter
in question for the luminosity function was the appropriate width along
the main sequence.  In theory the best extraction width might vary as a
function of magnitude since the photometric errors increase toward the
bottom of the CMD and since the main sequence slope changes with
luminosity.  In practice, because the comparison and cluster fields were
closely matched in the properties of the field stars and the photometric
depth, the width of the extraction window did not matter, at least within
a reasonable range.  The insensitivity of the total star counts in the LF
(from 19.82 $\leq V \leq$ 24.62) versus the extraction window width
(Figure 4) demonstrates this point.  Once the window width reached 0.29
mag ($\pm 0.14$ mag around the fiducial sequence) the total number of
stars in the LF remained constant, at 47.  In order to test for any
signature due to equal mass binary main sequence stars at $0.75$ mag above
the main sequence, we chose an extraction width of 0.15 mag ($\pm 0.07$
mag) around a locus offset by $-0.75$ mag.  This extraction width kept the
equal mass binary window from overlapping the main sequence extraction
window.  One untreated systematic in this procedure was the effect of
removing real cluster stars by using a conservative stellarity index.
Fortunately, as Figure 1 illustrates, real cluster stars may be missed
only for $I \le 22$, which is beyond the limit of the main sequence
luminosity function we extract (to $V = 24.62$).

As a final nuance to the derived cluster LF, we identify a small number of
objects in the control field as possible cluster stars, consistent with
star counts from the wide-field 0.9m photometry (derived from the data of
Deliyannis \etal 2002) at brighter magnitudes ($V \geq 18$) where we find
the cluster density in the control field may not diminish altogether to
zero, but rather appears to have dropped to $10 \pm 10$\% of the central
value.  This possible small contamination, with a significance of only
$1\sigma$, does not include systematic effects such as mass segregation,
which would increase the relative numbers of the faintest stars at greater
cluster radii.  Our control field thus may oversubtract cluster stars by
$\sim$10\% and our LFs may therefore slightly underestimate the true
cluster LF in this central field.  Since we are not making detailed
comparisons among our derived LF and the LFs of other clusters, or
attempting to recover M35's IMF, this possible and slight underestimate of
M35's LF does not pose a problem in our analysis.  Instead, we use M35's
LF to demonstrate the reality of our cluster fiducial sequence by showing
that we have actually found cluster stars along this sequence and to
demonstrate that the cluster contains very low mass stars to the limit of
our photometry, just slightly above the brown dwarf regime.

The differential M35 LF, binned in 0.5 mag intervals, is presented in
Figure 12a, along with error bars derived from the Poisson counting
statistics of the cluster and comparison field.  The presence of cluster
stars is most convincing at the faint end, which fortunately allows us to
define the fiducial sequence and compare it to stellar models, above.
Figure 12b presents the cumulative luminosity function for M35 along with
the cumulative LF for the cluster equal mass binaries.  The fraction of
binaries among the low mass stars in this field is low or even zero, even
though binaries are abundant ($\sim35$\%, Sung \& Bessell 1999) among
the high mass cluster stars.  The number of cluster stars continues to
increase to the LF limit, at $V = 24.6$.  In addition, Figure 12b also
presents the mapping between $V$-band magnitude and mass in solar units at
[Fe/H] = 0.0 for our assumed cluster distance and each of the stellar
isochrones discussed above.  This is presented in lieu of mass functions
for each stellar isochrone set, which would anyway have too few stars to
yield much information.  Clearly the different theoretical mass-luminosity
relations, at least in $M_{V}$, differ by too much for reliable mass
estimates in this mass range.

The low number of stars in our field and our focus on the fiducial main
sequence rather than the LF meant that a detailed completeness study to
correct the cluster LF was not warranted.  Nonetheless, we are able to
reliably and conservatively estimate completeness from the signal-to-noise
as a function of depth derived from past artificial star experiments with
nearly identical data taken at WIYN with the same detector and filters,
under similar and very low levels of crowding and sky illumination, but
under poorer seeing conditions (WOCS1).  Based on these earlier tests,
we estimate that the cluster LF is 90\% complete at $V = 24.4$, just before
the faint end of the extracted LF.  For comparison, 0.4 mag beyond the end
of the extracted LF, but nearly a magnitude above the bottom of the $VI$
CMD, at $V = 25.2$, we estimate completeness to be 50\%.

\subsection{The Observed Lower Mass Limit}

M35 contains stars to the limit of the extracted main sequence, at M
$\approx 0.10$--0.15 M$_{\odot}$ (using masses derived from the Baraffe
\etal 1998 models).  This suggests that M35 may harbor a large number of
brown dwarfs.  Where does the brown dwarf region in M35 start?  As
demonstrated in Figure 12b, estimating this magnitude depends on which
stellar isochrone set one adopts.  As an example, we adopt the results of
Chabrier, Baraffe, \& Plez (1996), who find that the H-burning minimum
mass for solar metallicity is $\sim0.070$ M$_{\odot}$.  (The slight
sub-solar metallicity for M35 ([Fe/H] = $-0.2$) should increase this
minimum mass by a small amount, $\sim0.002$ M$_{\odot}$ (Cabrier \etal
1996), though we ignore this small offset in calculating the H-burning
limit.)  Interpolating values from table 1 of Chabrier \etal for the
cluster age of 150 Myr yields the following properties for an H-burning
minimum mass object: $T_{\rm eff} = 2660$ K, log($L/L_{\odot}) = -3.02$,
$M_{V}$ = 15.8, $M_{I} = 12.6$, and $M_{K}$ = 9.3.  With the $V$-band
cluster distance modulus of 10.15 and E($B-V$) = 0.2 this is just beyond
the limit of our observations, at $V \approx 26$, corresponding to $I
\approx 22.7$ and $K \approx 19.1$.  Particularly in the near-IR such
observations are now readily obtainable even on 4m class telescopes.

\section{White Dwarfs}

Reimers \& Koester (1988a,b) obtained spectra of candidate white dwarfs
(WDs) identified in M35 by Romanishin \& Angel (1981) in a series of
papers meant to ascertain the upper mass limit for the formation of WDs
and to constrain the WD initial-final mass relation.  With an age of
$\sim150$ Myr, M35 has a turn-off mass of $\sim3.75$ M$_{\odot}$ (\eg from
the overshooting models of Girardi \etal 2000), and is certainly expected
to have produced white dwarfs.  Reimers \& Koester identified two objects,
which they call N2168-3 and N2168-4, as white dwarfs that are likely
cluster members with $T_{\rm eff}$ = 37500 and 44000 K, both with masses =
$0.7 \pm 0.1$ M$_{\odot}$, and cooling ages of $\sim1.5 \times 10^7$ yr.
They reported $V$-band magnitudes of 20.24 and 20.05 for these two
objects.  Their cooling ages and photometry are consistent with objects
which left the cluster main sequence when the cluster was $\sim80$\% its
current age.  Their mass estimate is perhaps $1\sigma$ lower than their
age derivation, more modern WD models, and the cluster age would imply.

While the Reimers \& Koester WDs were not in our field, we have found a
new candidate cluster WD at RA = $06^{h}09^{m}06^{s}.2$ and Dec =
$24^{\arcdeg}19^{\arcmin}25^{\arcsec}$ (J2000).  This candidate, along
with three other blue objects from the control field, can be found in
figures 2a through 2d near $B-V \approx V-I \approx 0.0$.  A closer look
at these four candidate WDs are presented in Figures 13a, 13b, and 14,
where they are compared to WD cooling tracks of Althaus \& Benvenuto
(1997, 1998) for H- and He-atmospheres and Wood (1992, and references
therein; with colors derived from interpolating in the tables of Bergeron,
Wesemael, \& Beauchamp 1995) for H-atmospheres, respectively.  In all
three figures the model cooling tracks with the highest and lowest masses
(in solar units) are labeled, as are illustrative ages (in Gyr) along the
tracks.  The listed ages are the total WD age, derived from the addition
of the WD cooling ages from the above-referenced models and our
determination of the precursor ages.  The precursor ages were derived by
converting WD masses used in the tracks to zero age main sequence masses
using the new initial-final mass relation of Weidemann (2000), along with
stellar evolutionary time scales up to the tip of the asymptotic giant
branch derived from interpolations of the Girardi \etal (2000) solar
metallicity models, including convective overshoot.  We chose the Girardi
\etal models because of their fine grid in mass and age for the range of
relevance here, though we checked them against time scales for models with
the same masses from Hurley, Pols, \& Tout (2000) and Yi \etal (2001), and
the differences in precursor ages were always $\leq 10$\%.  A 0.7
M$_{\odot}$ WD would have been a 3.17 M$_{\odot}$ star on the main
sequence with a pre-WD lifetime of 0.410 Gyr.  A 1.0 M$_{\odot}$ WD would
have been a 6.71 M$_{\odot}$ star on the main sequence with a pre-WD
lifetime of $< 0.063$ Gyr.  Uncertainties in these short precursor ages of
$\leq 10$\% have little effect ($< 6$ Myr) on the derived age for the
highest mass WDs expected to be the oldest in the cluster.

The blue objects from the control field are highly unlikely to belong to
the cluster due to both the very low number of cluster stars at this
location and due to the implied ages of $\geq 0.6$ Gyr.  These objects may
be field WDs, in which case they are most likely in the background and
therefore belong on lower mass tracks and are older, or they may be
background quasars.  The single blue object in the M35 central field with
$V = 21.36 \pm 0.01$ is a good candidate WD, based on the rarity of other
stars and blue compact galaxies at this magnitude, although its exact
interpretation if it is a cluster WD is unclear.  As a cluster WD,
according to the H-atmosphere cooling tracks of Althaus \& Benvenuto
(Figure 13a), it would have a mass of $1.05 \pm 0.05$ M$_{\odot}$ and an
age of 0.21--0.26 Gyr ($\pm 1 \sigma$ range), which is certainly possible,
but the lower limit is formally more than $4 \sigma$ higher than the
cluster age derived by Deliyannis \etal (2002) and WDs this massive are
rare.  The He-atmosphere cooling tracks of Althaus \& Benvenuto (Figure
13b) decrease the age slightly, to 0.19--0.22 Gyr, bringing them into
closer consistency (but still formally off by $3 \sigma$) with the
isochrone age, although the mass does not change appreciably.  Although
WDs this massive are rare in the field ($\sim1$\%), they are not as rare
in young clusters and the apparent high mass of the WD candidate is
sensible for the cluster age and turn-off mass.  According to the
Weidemann (2000) initial-final mass relation, this object would have been
a $\sim7$ M$_{\odot}$ star on the main sequence, making it potentially
important in the determination of the upper mass limit for WD creation.
On the other hand, interpreting this object with the H-atmosphere cooling
tracks of Wood (1992), yields a WD mass of 0.67--0.78 M$_{\odot}$ and an
age of 0.42--0.83 Gyr.  We do not know the source of the inconsistency
between Althaus \& Benvenuto on the one hand and Wood and Bergeron \etal
(the source of the temperature-color calibration) on the other hand.  To
further test the WD models, as well as determine if the candidate cluster
WD is a {\it bona fide} cluster WD, follow-up spectroscopy is required.
If the object turns out to be a cluster WD then the nature of its
atmosphere (H or He) as well as its surface gravity (if an H-atmosphere
WD) will yield its mass and further test the WD models, while at the same
time intercalibrating the WD and main sequence models.  The differences in
the derived main sequence and WD ages could be due to real problems in
stellar evolution of intermediate mass stars, where age depends
sensitively on the degree of mixing in convective cores.  Accurate
parameters for this star and data able to discern its possible cluster
membership would be particularly valuable.

\section{Conclusion}

We obtained deep $BVI$ observations of M35 and a nearby comparison field
with the WIYN 3.5m telescope under non-photometric but excellent seeing
conditions.  We calibrated the data against shallower 0.9m data (from
Deliyannis \etal 2002), achieving a photometric accuracy of approximately
0.02 mag.  These deep observations display the lower main sequence in the
$BV$ and $VI$ CMDs down to $V$ = 23.3 and 24.6, respectively.  At these
faint magnitudes the background Galactic field stars are far more numerous
than the cluster stars, yet by using a smoothing technique (Silverman
1986) and CMD density distribution subtraction we were able to recover the
cluster fiducial main sequence and luminosity function to $V = 24.6$.  We
find the location of the M35 main sequence in these CMDs to be consistent
with earlier work on other open clusters, specifically NGC 188 (WOCS1),
NGC 2420 (von Hippel \& Gilmore 2000), and NGC 2477 (von Hippel \etal
1996).  On comparing these open cluster fiducial sequences to stellar
models by Baraffe \etal (1998), Siess \etal (2000), Girardi \etal (2000),
and Yi \etal (2001) we find that the models are too blue in both $B-V$ and
$V-I$ for stars less massive than $\sim0.4$ M$_{\odot}$.  At least part of
the problem appears to be underestimated opacity in the bluer bandpasses,
with the amount of missing opacity increasing toward the blue.

M35 contains stars to the limit of the extracted main sequence, at M
$\approx$ 0.10--0.15 M$_{\odot}$, suggesting that M35 may harbor a large
number of brown dwarfs.  These brown dwarfs should be easy targets for
sensitive near-IR instrumentation now being mounted on 8--10m telescopes.
In fact, imaging observations of only one hour in $K$ would allow one to
obtain S/N = 30 photometry 1 magnitude fainter than the brown dwarf limit
in this cluster.

We also identify a new candidate white dwarf in M35 at $V = 21.36 \pm
0.01$.  Depending on which WD models are used in interpreting this
object, it is either a very high mass WD ($1.05 \pm 0.05$ M$_{\odot}$)
somewhat (3--$4\sigma$) older than our best isochrone age (150 Myr), or it
is a modestly massive WD (0.67--0.78 M$_{\odot}$) much too old (0.42--0.83
Gyr) to belong to the cluster.  Follow-up spectroscopy is required to
resolve this issue.

\acknowledgments

T. von Hippel would like to thank Isabelle Baraffe, Pierre Demarque, Karl
Gebhardt, and Don Winget for helpful discussions; Isabelle Baraffe, Gilles
Chabrier, France Allard, and Peter Hauschildt for their unpublished
models; and Karl Gebhardt for the software used for the field star
subtraction procedure.  We also thank an anonymous referee for a number
of suggestions that improved the rigor and clarity of the paper.  A.
Sarajedini and T. von Hippel gratefully acknowledge support for this work
from NSF grants AST-9819768 and AST-0196212, and C. P. Deliyannis
gratefully acknowledges support from NSF grant AST-9812735.

\clearpage

\figcaption{$I$-band magnitude versus stellarity index for the M35 central
field (solid circles) and the control field (open circles).  The control
field points have been shifted down by 0.1 and to the right by 1.0 mag for
visibility.  The stellarity index cut of 0.95, the threshold for inclusion
of the objects in the CMDs and subsequent analyses, is indicated for both
fields.}

\figcaption{$B-V$ versus $V$ CMD for ({\it a}) the cluster central field
and ({\it b}) the control field.  Typical error bars for cluster main 
sequence stars are plotted along the right.  The fiducial main sequence
is also indicated by the solid line.  $V-I$ versus $V$ CMD for ({\it c})
the cluster central field and ({\it d}) the control field.}

\figcaption{The smoothed version of the $VI$ CMD for the ({\it a}) cluster
central field and ({\it b}) the comparison field with a smoothing kernel
of 0.10 mag.  The horizontal axes are $V-I$ color from 0.0 to 4.5 and the
vertical axes are $V$ magnitude from 15 to 26.  The difference CMD ({\it
c}) exhibits a subtle but clear main sequence just above the solid white
line, itself the fitted fiducial offset 0.3 mag fainter for display
purposes.  Subtraction noise throughout the field star region is evident,
though the net star number there is close to zero.}


\figcaption{The cumulative star count in the extracted $VI$ luminosity
function as a function of the width of the $V-I$ extraction window.  The
solid line running through open circles presents the cumulative star count
for an extraction window centered on our initial cluster fiducial
sequence.  From left to right the curves present cumulative star counts
for windows shifted by $V-I = +0.025, 0.000, +0.050, -0.025, +0.075,
-0.050, +0.100, -0.075, +0.125, -0.100,$ and $-0.125$ mag, where positive
shifts are toward the red.  The fiducial offset by $V-I = +0.025$ is the
one we adopt as the best fit.  After a window width of 0.29 mag ($\pm
0.14$ mag around the fiducial sequence) the number of stars remains
essentially constant for our chosen and nearby fiducial sequences,
demonstrating the stability of the CMD subtraction technique.}

\figcaption{Comparison of the upper portion of the fiducial sequence to the
Hipparcos-based fiducial sequence derived by Pinsonneault \etal (1998) in
the dereddened $VI$ CMD, indicating good consistency with our adopted
cluster distance and reddening.  The dashed line extends across the region
of the Pinsonneault \etal calibration and the error bars are estimated
uncertainties in the individual fiducial points.}

\figcaption{Comparison between our M35 fiducial main sequence and the adopted
main sequence of Barrado y Navascu\'es \etal (2001b), itself a combination
of bright M35 stars from Sung \& Bessell (1999) and faint solar
neighborhood stars from Leggett (1992).}

\figcaption{Comparison between the M35 fiducial main sequence and the
fiducial main sequences of the older open clusters NGC 188, NGC 2420, and
NGC 2477.  The portion of the M35 fiducial derived from the Kitt Peak 0.9m
and WIYN data is indicated by a dotted line and a solid line connecting
the filled circle symbols, respectively.  NGC 188 is plotted as open circles
along a solid line.}

\figcaption{({\it a}) Comparison between the M35 and NGC 188 main
sequences and the solar metallicity models of Baraffe \etal (1998).
Stellar models for 158 Myr and 6.3 Gyr are indicated, as are a few
representative mass values in solar units along the 158 Myr isochrone.
({\it b}) Similar to {\it a} but the Baraffe \etal models employ an
updated TiO line list from Schwenke (1998).}

\figcaption{Comparison between the M35 main sequence in the ({\it a})
dereddened $BV$ CMD with Siess \etal (2000) isochrones, using either the
color transformations of Siess \etal (1997, full line) or Kenyon \&
Hartmann (1995, dashed line).  The same comparison in the ({\it b})
dereddened $VI$ CMD, now including NGC 188 and a comparably aged set of
isochrones.}

\figcaption{Similar to Figures 9a and 9b, but now comparing the cluster
fiducials to solar metallicity Girardi \etal (2000) isochrones.}

\figcaption{Similar to Figures 9a and 9b, but now comparing the cluster
fiducials to Yi \etal (2001) isochrones.  For both clusters two isochrones
are plotted for two different color transformations.  The dotted line
indicates the transformation derived by Green \etal (1987) and the solid
line the transformation derived by Lejeune \etal (1998).}

\figcaption{The ({\it a}) differential and ({\it b}) cumulative luminosity
function extracted from the $VI$ CMD.  The differential luminosity
function is plotted with error bars derived from the Poisson statistics of
the subtraction process.  The cumulative LF for both the main sequence
stars and any possible equal mass binary signature are plotted, along with
mass values from various models.  The symbols to the left of the mass
values indicate their origin, and are B = Baraffe \etal (1998), S$_{\rm
K}$ = Siess \etal (2000) employing the Kenyon \& Hartmann (1995) color
transformation, S = Siess \etal (2000) employing the Siess \etal (1997)
color transformation, G = Girardi \etal (2000), and Y = Yi \etal (2001).
Incompleteness corrections, which amount to $\sim 10$\% at the faint end
of the LF, are not included.}

\figcaption{The candidate cluster WD (square symbol) from the central
field and candidate field WDs (circle symbols) from the control field
compared to ({\it a}) H-atmosphere and ({\it b}) He-atmosphere WD cooling
tracks from Althaus \& Benvenuto (1997, 1998).  The highest and lowest
mass cooling tracks are marked with their mass in solar units.  Total WD
ages in Gyr along some of the cooling tracks are also indicated.}

\figcaption{Same as Figure 13a, but for Wood (1992) cooling tracks,
incorporating the $T_{\rm eff}$-color transformations of Bergeron \etal
(1995).}

%
\begin{deluxetable}{lccccccrc}
\tablewidth{0pt}
\tablecaption{Calibration to Landolt System}
\tablehead{
\colhead{filter} & \colhead{frames}  & \colhead{stars} & \colhead{zeropt} & 
\colhead{err}    & \colhead{airmass} & \colhead{err}   & \colhead{color}  & \colhead{err} \\
\phantom{1234}(1) & (2) & (3) & (4) & (5) & (6) & (7) & (8)\phantom{12} & (9)}
\startdata
$B$          & 7 & 63 & 3.4567 & 0.0149 & 0.2235 & 0.0114 & $-0.0814$ & 0.0035 \\
$V$=f($B-V$) & 7 & 65 & 3.2405 & 0.0134 & 0.1179 & 0.0102 &   0.0107  & 0.0031 \\
$V$=f($V-I$) & 7 & 67 & 3.2413 & 0.0137 & 0.1167 & 0.0104 &   0.0111  & 0.0029 \\
$I$          & 7 & 77 & 3.8970 & 0.0121 & 0.0448 & 0.0090 & $-0.0122$ & 0.0024
\enddata
\end{deluxetable}

%
\begin{deluxetable}{ll}
\tablewidth{0pt}
\tablecaption{$BV$ Fiducial Main Sequence}
\tablehead{
\colhead{$B-V$} & \colhead{$V$} \\
\phantom{1}(1) & \phantom{1}(2)}
\startdata
%
0.15 & 11.31 \\ 
0.2  & 11.72 \\ 
0.3  & 12.28 \\ 
0.4  & 12.78 \\ 
0.5  & 13.26 \\ 
0.6  & 13.80 \\ 
0.7  & 14.42 \\ 
0.8  & 15.06 \\ 
0.9  & 15.61 \\ 
1.0  & 16.00 \\ 
1.1  & 16.52 \\ 
1.19 & 16.85 \\ 
1.31 & 17.3  \\ 
1.35 & 17.5  \\ 
1.55 & 18.4  \\ 
1.71 & 19.1  \\ 
1.77 & 19.4  \\ 
1.83 & 19.82 \\ 
1.85 & 20.12 \\ 
1.89 & 20.48 \\ 
1.91 & 20.96 \\ 
1.93 & 21.24 \\ 
1.93 & 21.40 \\ 
1.93 & 21.70 \\ 
1.99 & 22.26 \\ 
2.03 & 22.86 \\ 
2.09 & 23.32 
\enddata
\end{deluxetable}

%
\begin{deluxetable}{ll}
\tablewidth{0pt}
\tablecaption{$VI$ Fiducial Main Sequence}
\tablehead{
\colhead{$V-I$} & \colhead{$V$} \\
\phantom{1}(1) & \phantom{1}(2)}
\startdata
%
0.3  & 11.7  \\
0.4  & 12.3  \\
0.5  & 12.8  \\
0.6  & 13.05 \\
0.7  & 13.4  \\
0.8  & 14.05 \\
0.9  & 14.7  \\
1.0  & 15.35 \\
1.1  & 15.95 \\
1.2  & 16.4  \\
1.3  & 16.9  \\
1.4  & 17.25 \\
1.5  & 17.5  \\
1.53 & 17.7  \\
1.82 & 18.5  \\
2.16 & 19.4  \\
2.42 & 19.95 \\
2.59 & 20.5  \\
2.75 & 20.96 \\
2.99 & 21.70 \\
3.11 & 22.26 \\
3.27 & 22.84 \\
3.39 & 23.46 \\
3.55 & 24.02 \\
3.67 & 24.34 \\
3.83 & 24.62 
\enddata
\end{deluxetable}

\end{document}